\documentclass[preprint2]{aastex62}

\usepackage{mathtools}
\usepackage{graphicx}

\shorttitle{GeV-TeV EBL Measurement}
\shortauthors{Desai et al.}
\usepackage{calrsfs,euscript,mathrsfs}
\usepackage[outdir=./]{epstopdf}

\begin{document}

\title{A GeV-TeV Measurement of the Extragalactic Background Light}

\author{A.~Desai}
\affiliation{Department of Physics and Astronomy, Clemson University, Kinard Lab of Physics, Clemson, SC 29634-0978, USA}
\affiliation{abhishd@g.clemson.edu}
\author{K.~Helgason}
\affiliation{Science Institute, University of Iceland, IS-107 Reykjavik, Iceland}
\affiliation{helgason@hi.is}
\author{M.~Ajello}
\affiliation{Department of Physics and Astronomy, Clemson University, Kinard Lab of Physics, Clemson, SC 29634-0978, USA}
\author{V.~Paliya}
\affiliation{Deutsches Elektronen Synchrotron DESY, Platanenallee 6, 15738 Zeuthen, Germany}
\author{A.~Dom\'\iınguez}
\affiliation{Grupo de Altas Energ{\'i}as and IPARCOS, Universidad Complutense de Madrid, E-28040 Madrid, Spain}
\author{J.~Finke}
\affiliation{Space Science Division, Naval Research Laboratory, Washington, DC 20375-5352, USA}
\author{D.~Hartmann}
\affiliation{Department of Physics and Astronomy, Clemson University, Kinard Lab of Physics, Clemson, SC 29634-0978, USA}



\begin{abstract}

The Extragalactic Background Light (EBL) can be probed via the absorption imprint it leaves in the spectra of gamma-ray sources ($\gamma\gamma \rightarrow e^-e^+$). 
We recently developed a dedicated technique to reconstruct the EBL, and its evolution with redshift, from $\gamma$ ray optical depth data using a large sample of blazars detected by the {\it Fermi} Large Area Telescope. Here, we extend this dataset to the TeV regime using ground-based Cherenkov observations of 38 blazars and report the first homogeneous measurement of the EBL spectral intensity covering the ultraviolet to infrared wavelengths ($\sim$0.1-100$\mathrm{\mu m}$). A minimal EBL throughout the wavelength range with respect to integrated galaxy light is found, allowing little additional unresolved emission from faint or truly diffuse populations setting an upper limit of $\lesssim 4~{\rm nW\cdot m^{-2}sr^{-1}}$ at 1.4\,${\rm \mu m}$. In particular, the cosmic optical background (COB) at $z=0$ is found to be $27.8_{-2.0}^{+2.1}~{\rm nW\cdot m^{-2}sr^{-1}}$. This work lays the foundation for accurate gamma-ray measurements of the EBL across its whole spectral range using a combination of GeV and TeV data. 

\end{abstract}


\accepted{02/27/19 \apjl}

%
%
\section{Introduction}

The Extragalactic Background Light (EBL) is the diffuse background radiation accumulated over the cosmic history at ultraviolet (UV), optical and  infrared (IR) wavelengths \citep[e.g.][]{dwek13}. The local EBL energy spectrum comprises  two peaks, with the first peak ($\approx$ 1 $\mu$m) due to direct emission from stars and the second peak ($\approx$ 100 $\mu$m) due to reprocessed star-light emission by dust within galaxies\citep{brun13}. Measurements of both the EBL spectral intensity and its evolution are important to study both star formation and galaxy evolution processes \citep[e.g.][]{raue12,Cowley2018,Khaire18}.

Measuring the EBL brightness has proven challenging mainly due to bright foreground contaminants such as the Zodiacal Light and the Diffuse Galactic Light \citep{Hauser98}. Studying the signatures left by the EBL in the spectra of distant $\gamma$ ray sources, via the photon-photon interaction, is emerging as the most powerful technique to probe the EBL. Various attempts have been made to constrain the EBL intensity using the absorption found in the spectrum of Blazars at { 0.1-100\,GeV} high energy (HE) and 0.1-30\,TeV very high energy (VHE). The constraints came first in the form of upper limits on the intensity \citep[e.g.][]{Aharonian06,mazin07,Meyer12} and later as measurements of the actual levels \citep[e.g.][]{ebl12,hess_ebl13,biteau2015,Ahnen16}. However, the majority of these measurements rely on scaling existing EBL models \citep[like the models of][]{kneiske10,finke10,dominguez11,gilmore12,stecker16} in amplitude. To improve this, we have developed a method to reconstruct the EBL spectrum and evolution based on measured $\gamma$ ray optical depths \citep[see ][]{Science2018}. 

In this Letter, we apply this newly developed tool to measure the EBL using both GeV and TeV data. While the GeV optical depths are taken from \cite{Science2018}, the TeV optical depths are derived using the multiple spectra of 38 TeV blazars reported in \cite{biteau2015}. The combined data-set enables us to consistently constrain the EBL spectral intensity in the wavelength regime 0.1\,$\mu$m to 100\,$\mu$m. The paper is organized as follows: in Section~\ref{sec:analysis} we describe the procedure used to derive the TeV optical depths, in Section~\ref{sec:EBL} we describe the methodology used to reconstruct the EBL and in Section~\ref{sec:discussion} we discuss the implications of our measurements.



%
%
\section{Analysis}
\label{sec:analysis}

\subsection{The intrinsic blazars' spectra}
\label{sec:intrinsicanalysis}

Our analysis relies on the 106 VHE gamma-ray spectral energy distributions (SEDs) of 38 blazars reported in \cite{biteau2015}. The source photons in this sample originate from $z=0.019$ to $z=0.604$ and are detected in the 0.1\,TeV to 21\,TeV range.
The SEDs are modeled in this energy range using:

\begin{equation}
    {\frac{dN}{dE}}_{\text{ obs}} = {\frac{dN}{dE}}_{\text{ int}} \cdot e^{-b \cdot \tau_{\text{ model} }}
    \label{eq1}
\end{equation}

\noindent where, ${{dN/dE}_{int}}$ and ${{dN/dE}_{obs}}$ are the intrinsic and observed blazar spectrum respectively, $\tau_{model}(E)$ is the optical depth estimated by EBL models at the source redshift \citep[e.g.][]{kneiske10,finke10,dominguez11,gilmore12} and $b$ is a renormalization constant to scale the optical depth.

To model the intrinsic spectrum we follow the methodology similar to { previous analyses} of VHE data where four different intrinsic spectral functions are used \citep[see also  ][]{biteau2015,acciari19}. These models are power law, log-parabola, power law with exponential cutoff and log-parabola with exponential cutoff. For a given EBL model (see e.g. Tab.~1), the intrinsic spectrum of a source is then chosen by adopting the function that produces the highest $\chi^2$ probability when $b=1$.

\begin{figure*}[ht!]
   \begin{center}
   \begin{tabular}{c}
    \includegraphics[scale=0.5,trim={0.6cm 0.6cm 1.5cm 1cm},clip]{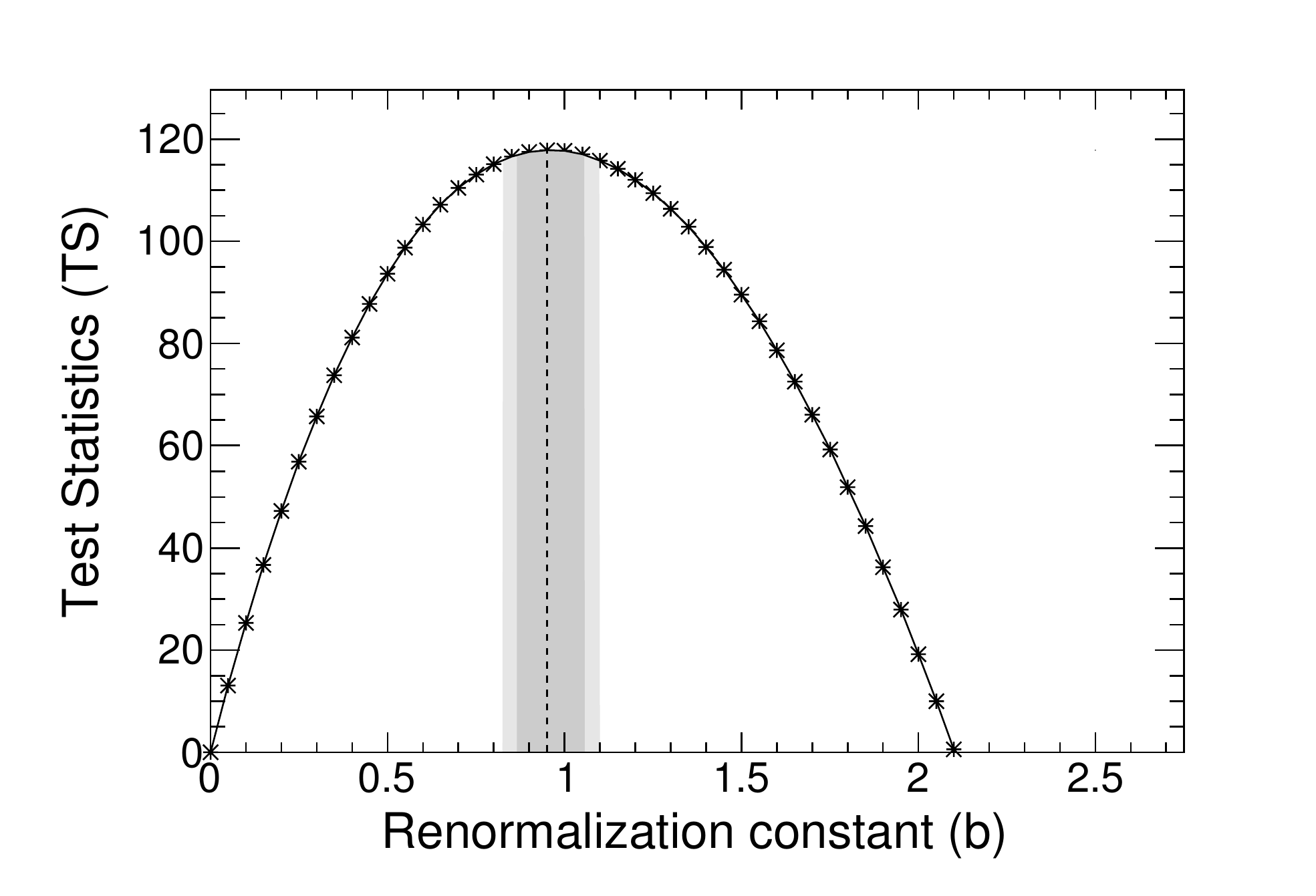}
    \includegraphics[scale=0.5,trim={0.6cm 0.6cm 1.5cm 1cm},clip]{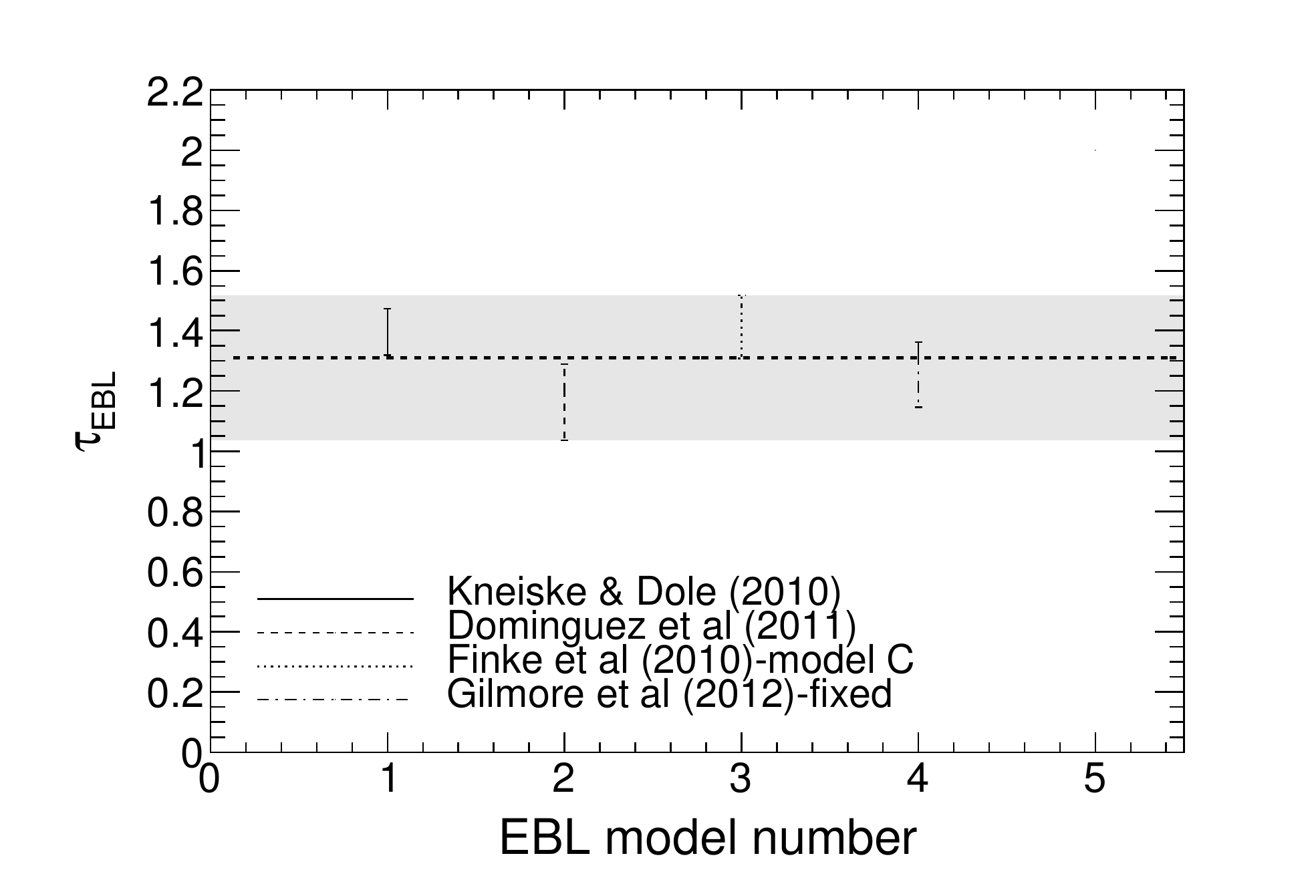}
   \end{tabular}
   \end{center}
   \caption{Left: Stacked TS profile obtained for a range of $b$ values using the EBL optical depths given by  \cite{biteau2015}. The dotted line shows the best-fit value of $b$ for the given EBL model. The shaded dark gray region highlights the 1\,$\sigma$ contour while light gray region shows the 2\,$\sigma$ contour. {Right: Example of how the optical depth is derived for a given energy and redshift bin using the values of $\tau$ derived using four different EBL model. The horizontal dashed line shows the best-fit value of the optical depth, while its uncertainty (the gray band) is chosen to encompass the uncertainty of all models.}
  \label{fig:ts_b}}
\end{figure*}

\begin{deluxetable}{lccc}

\tablewidth{0pt}
\tablecaption{Results of EBL models tested using VHE data}
\tablehead{\colhead{Model} &
$b_{best-fit}$  \tablenotemark{a}&
Test Statistic  \tablenotemark{b}
}
\startdata 
{\it \cite{finke10} --  model C}& $1.05^{-0.15}_{+0.12}$ & 71.60 \\ 
{\it \cite{kneiske10}} & $1.4^{-0.16}_{+0.14}$ & 99.47\\ 
{\it \cite{dominguez11}} & $0.85^{-0.10}_{+0.08}$ & 105.98\\ 
{\it \cite{gilmore12}  -- fixed} & $1.00^{-0.16}_{+0.12}$ & 106.13 \\ 
{\it \cite{biteau2015}} & $0.95^{-0.08}_{+0.11}$ & 117.83 \\
\enddata
\tablenotetext{a}{Best-fit renormalization constant derived from the stacking analysis}
\tablenotetext{b}{TS obtained by comparing the log likelihood obtained for the null case of $b=0$ with the value obtained for $b=b_{best-fit}$.} 
%
\label{tab:result}
\end{deluxetable}

For a given EBL model and for each source, a likelihood profile of the re-normalization constant  $b$ is produced. This is transformed into a { test statistic} (TS) profile, by subtracting the value of the log-Likelihood at $b$=0 and multiplying by two. For a given EBL model, the TS profiles of all sources are summed generating a
``stacked" TS, which allows us to identify the best-fit value of $b$ { for all spectra}.
%
This stacked TS value also displays the significance of the result as $\sqrt{TS} $ \citep[see also][]{ebl12,hess_ebl13,desai17}. 
The stacked TS vs $b$ profile for one EBL model is shown in Fig ~\ref{fig:ts_b}. There is a maximum TS of 117.83, which implies a detection at approximately $10.85\,\sigma$. Table~\ref{tab:result} shows the results obtained using five different EBL models. 



\subsection{Deriving the EBL optical depths}
\label{sec:opdepth}

In order to measure the optical depth, we perform a stacking analysis where the source sample is divided into two redshift bins and the analysis is performed across four energy bins. While the two redshift bins ($0.01<z<0.04$ and $0.04<z<0.604$) are chosen such that they contain the same signal strength (TS contribution to the analysis described in Section ~\ref{sec:intrinsicanalysis}), the energy bins are chosen to have equal logarithmic widths.
For each energy and redshift bin, a stacked TS vs $b$ profile is derived using the method described in Section ~\ref{sec:intrinsicanalysis} where the source sample and energy range is modified according to the bin being considered. The corresponding $b$ value, in each redshift and energy bin, is then used alongside the EBL model being tested to obtain the optical depth (as obtained using that model).

\begin{figure*}[ht!]
  \begin{center}
  \begin{tabular}{c}
    \includegraphics[scale=0.5]{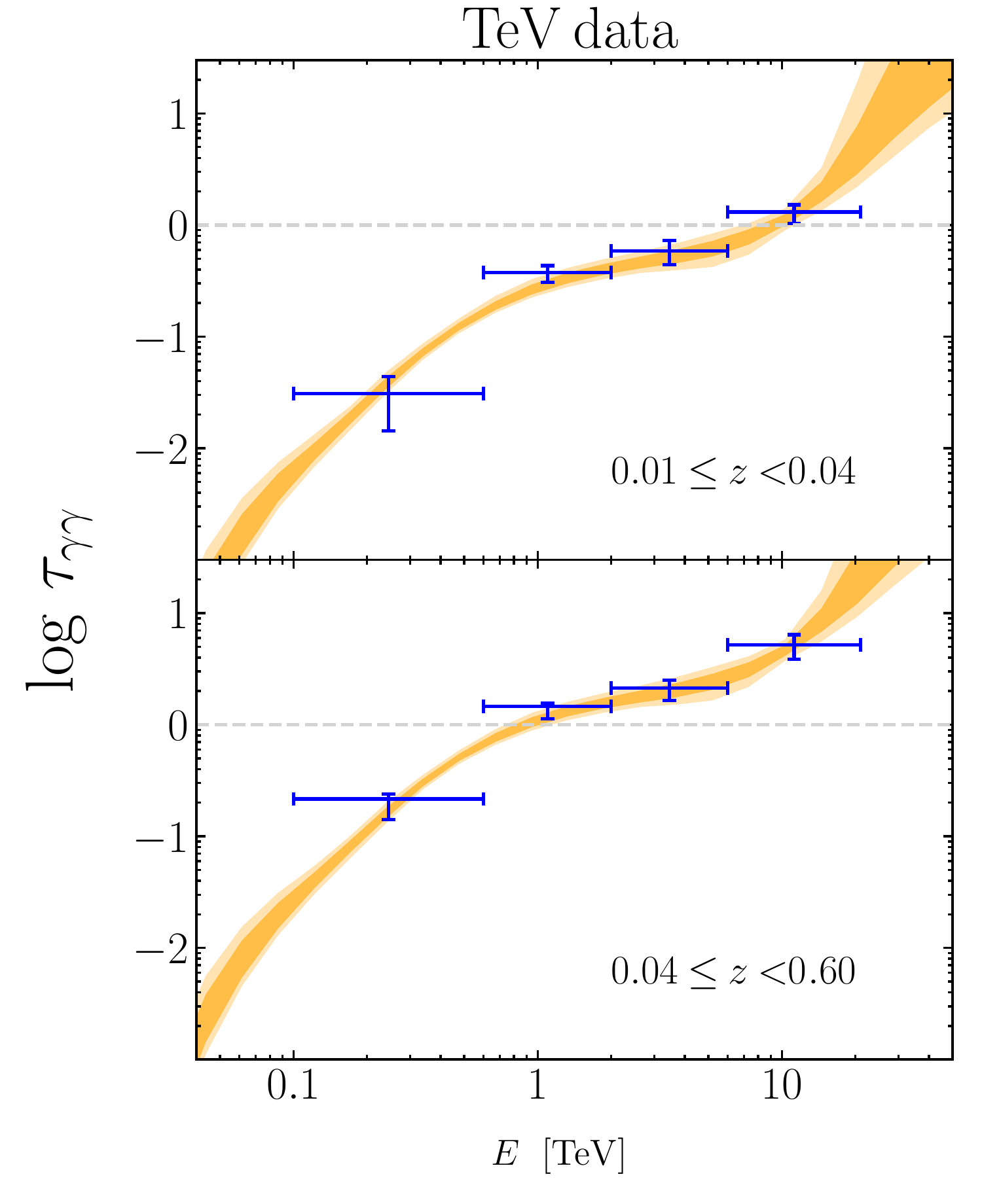}
    \includegraphics[scale=0.5]{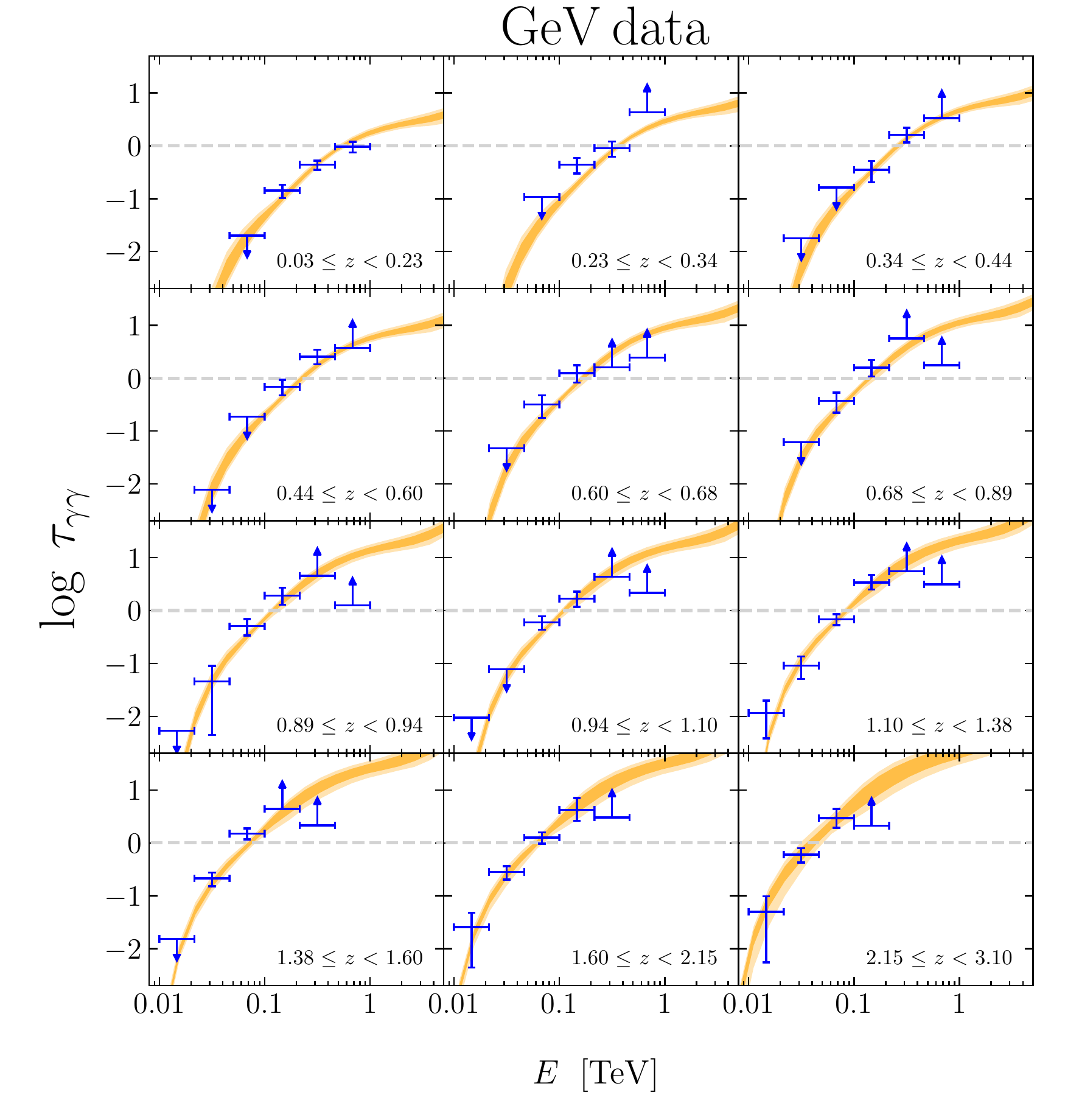}
  \end{tabular}
  \end{center}
  \caption{Redshift binned optical depth measurements derived from the stacking analysis using VHE data (left) and {\it Fermi}-LAT data (right) are shown compared to the optical depth templates reported by this work. The shaded regions signify the $1 \sigma$ and $2 \sigma$ confidence regions of our best-fitting EBL reconstruction.
\label{fig:opdep}}
\end{figure*}

We perform the above binned analysis for four\footnote{The optical depths reported in \cite{biteau2015} were optimized relying on the data used in this work and as such are not used here.} different EBL models \citep{finke10,kneiske10,dominguez11,gilmore12}.
{ In a given energy and redshift bin, the optical depth is derived as the mean of the  four individual optical depth measurements (derived using four different EBL models), while the uncertainty is chosen as the one that encompasses the uncertainties of all the optical depth measurements, as shown in Figure ~\ref{fig:ts_b}.} Along with the statistical uncertainties, the uncertainty on the optical depth includes a systematic contribution due to the difference in shape of the optical depth curve estimated from EBL models, the intrinsic model used in Equation ~\ref{eq1} and the systematic energy bias of $\approx10\%$ found in TeV data measured using Cherenkov telescopes\citep{meyer10}. As in \cite{biteau2015}, the impact of these systematic uncertainties on the EBL optical depth is estimated to be $\approx 2-5\%$. We include these uncertainties in our measurement of the derived optical depth values and show it in Fig.~\ref{fig:opdep}. These optical depth measurements are also made available on an online database.\footnote{\url{https://figshare.com/s/9cd4f26925945470582a}. }

%
%
\section{Reconstructing the EBL}
\label{sec:EBL}

\begin{figure*}
  \begin{center}
    \includegraphics[scale=0.6]{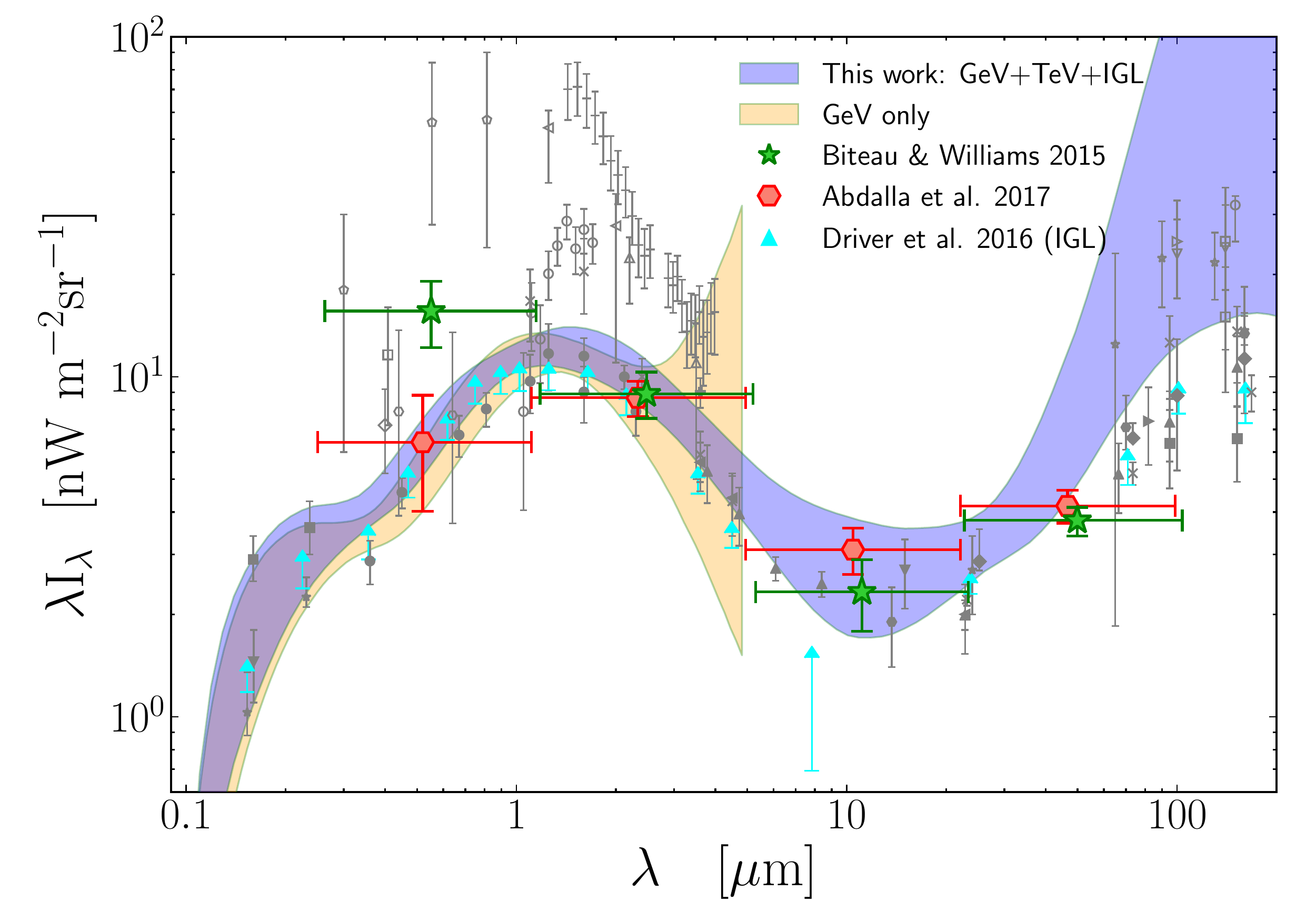}
  \end{center}
  \caption{ The spectral intensity of the EBL from UV to far-IR. The constraints from this work are shown as a 68\% confidence region and median (blue). A corresponding region from \citep{Science2018} that relies on GeV data only is shown in orange. Various measurements in the literature are shown in gray: direct measurements (open symbols), integrated galaxy counts (filled symbols). The numerical data of the blue and orange curves are available at 
  \url{https://figshare.com/s/9cd4f26925945470582a}.
\label{fig:ebl}}
\end{figure*}
We use the derived optical depths to reconstruct the absorbing EBL in a model independent way. In what follows, we include optical depth measurements at $\lesssim 1$ TeV by \cite{Science2018} based on a sample of 739 blazars observed by {\it Fermi}-Large Area Telescope (LAT). These measurements provide $\tau(E,z)$ in twelve redshift bins in the $z=0.03-3.1$ range and are thus highly complementary to our VHE data-set. Whereas the optical depths at VHE constrain the local EBL from optical to far-IR wavelengths, the {\it Fermi}-LAT data-set probes the UV to optical out to high redshifts. In terms of wavelength and redshift coverage, the combined data-set results in the most extensive constraints of the EBL to date.

We follow the novel methodology presented by \cite{Science2018}, where the cosmic emissivity (luminosity density) is modeled as the sum of several log-normal templates with a fixed peak position

\begin{equation} 
\label{eqn:jsum}
j(\lambda) = \sum_i a_i \cdot \exp \left[ -\frac{\left (\log \lambda -
      \log \lambda_i \right)^2 }{2 \sigma^2}   \right]
\end{equation}

\noindent in ${\rm erg\cdot s^{-1}cm^{-3}{\AA}{\rm ^{-1}}}$ where the pivots are logarithmically spaced with $\Delta_{\rm lg\lambda}=0.5$, resulting in seven templates centered at $\lambda_i = [0.16,0.50,1.6,5.0,16,50,160] {\rm \mu m}$. 
We fix $\sigma=0.2$, and leave the amplitudes $a_i$ free to vary. 
We tried varying the number of templates and their placement under the condition that $\sigma = \Delta_{\rm lg\lambda}/2.5$ and find that the local EBL is always consistent within the one sigma confidence region of the final result shown in Figure \ref{fig:ebl}. 
Each template is allowed to evolve independently with redshift according to

\begin{equation} \label{eqn_jnu}
  j(\lambda_i,z) = 
  j_0(\lambda_i) \cdot 
  \begin{cases*}
    \frac{(1+z)^{b_i}}{1+\left( \frac{1+z}{c_i}\right)^{d_i}}, & $i\leq 3$ \\
(1+z)^{b_i}, & $i>3,$ 
\end{cases*}
\end{equation}

\noindent where $j_0(\lambda_i) \equiv j(\lambda_i,z=0)$ is the emissivity at the present time centered at $\lambda_i$. We reduce the number of parameters by splitting the evolution form at $\lambda \simeq 5{\rm \mu m}$ as the TeV sources are only sensitive to EBL photons at low-$z$ towards infrared wavelengths. This results in 18  free parameters.

The local EBL is obtained from the evolving emissivity $j(\lambda,z)$:
\begin{equation}
 \lambda I_\lambda = \frac{c}{4\pi} \int \lambda^\prime j(\lambda^\prime,z) \frac{dt}{dz}\frac{dz}{(1+z)}
\end{equation}
where $\lambda^\prime = \lambda/(1+z)$ is the rest-frame wavelength.

The MCMC code \texttt{emcee} \citep{Foreman13}, a Python implementation of an affine invariant MCMC ensemble sampler \citep{GoodmanWeare10}, is used to constrain the parameters controlling the emissivity. With the emissivity specified as a function of wavelength and redshift, we calculate the resulting EBL and optical depth at 14 redshifts corresponding to each of the bins in Figure \ref{fig:opdep}. In addition to the optical depth data, we have included the integrated galaxy counts from \citet{driver16} as lower limits on the EBL at $z=0$. These are taken to be the lower uncertainty for the galaxy light data obtained by integrating over the observed magnitude range only i.e. not extrapolated\footnote{We take the lower of the two PACS160 values given in Table 2 of \citet{driver16}}. The likelihood function is estimated as ${\mathcal L} \propto \exp{(-\chi^2)}$ where the total number of optical depth data points and EBL lower limits used to calculate $\chi^2$ is 97. Our final results are based on MCMC chains from 120 walkers exploring the parameter space in 10,000 steps each. This results in 1,140,000 steps after a burn-in of 500 steps for each walker. We refer to \cite{Science2018} for details.

%
%

\section{Discussion}
\label{sec:discussion}

The measured constraints on the local EBL with a 68\% confidence region are displayed in Figure \ref{fig:ebl}. For a comparison, we also show previous measurements reported in the literature. { While the  results  are  in  good  agreement  with  the {\it Fermi}-LAT  measurement  (orange)  relying  on GeV data only \citep{Science2018}, a minor difference is seen in the higher end of the uncertainty at $\approx2\,\mu m$ which is mainly driven by the Cherenkov measurements of the repeatedly observed Mkn 421 and Mkn 501.}

{ The slightly larger-intensity of the GeV$+$TeV$+$IGL result, as compared to GeV only, is not entirely due to the inclusion of the IGL lower limits. In fact, examining the reduced $\chi^2 = \chi^2_{\rm GeV+TeV} + \chi^2_{\rm IGL}$ shows that the GeV$+$TeV dataset prefers this intensity independently of the IGL lower limits. In other words, lowering the EBL intensity does not improve $\chi^2_{\rm GeV+TeV}$. This is also reflected in the reconstructed optical depths shown in Figure \ref{fig:opdep} which do not show systematically higher optical depths with respect to the data. However, we find that the IGL lower limits help constrain the spectral shape of the EBL, making the result less dependent on the placement of the spectral templates ($\lambda_i$). This is not surprising since the optical depth at a given energy is an integral over the EBL wavelengths encompassed by the photon-photon interaction cross-section.}

An overall agreement is found with independent EBL measurements, both integrated galaxy counts and other $\gamma$-ray absorption studies. The combined GeV$+$TeV data-set is also sensitive to EBL photons in the mid-IR (${\rm \lesssim 100 \mu m}$) where we find good agreement with previous studies \citep{biteau2015,driver16,hess17}. However, the $\gamma$ ray dataset has no constraining power at ${\rm \gtrsim 100 \mu m}$ and the lower limits therefore push the EBL to higher far-IR values. 

Our measurements are particularly valuable in the UV/optical, where previous $\gamma$-ray absorption studies had limited sensitivity , integrated counts show conflicting results \citep{Gardner00,Xu05,Voyer11} and direct measurements remain somewhat above the counts data \citep{Bernstein07,matsuoka11,mattila17}.

\begin{figure*}[ht!]
  \begin{center}
  \includegraphics[width=0.39\textwidth]{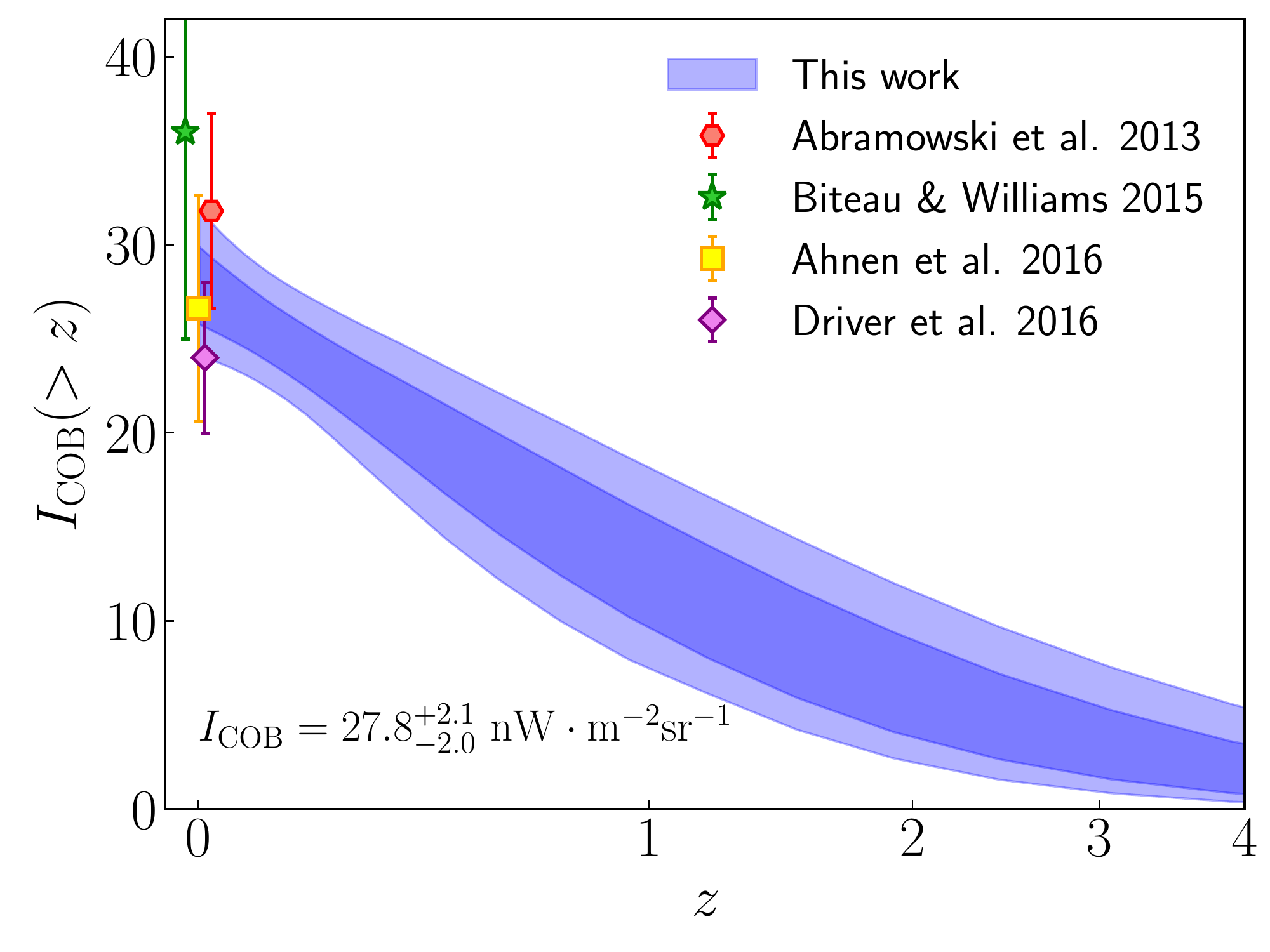}
  \includegraphics[width=0.6\textwidth]{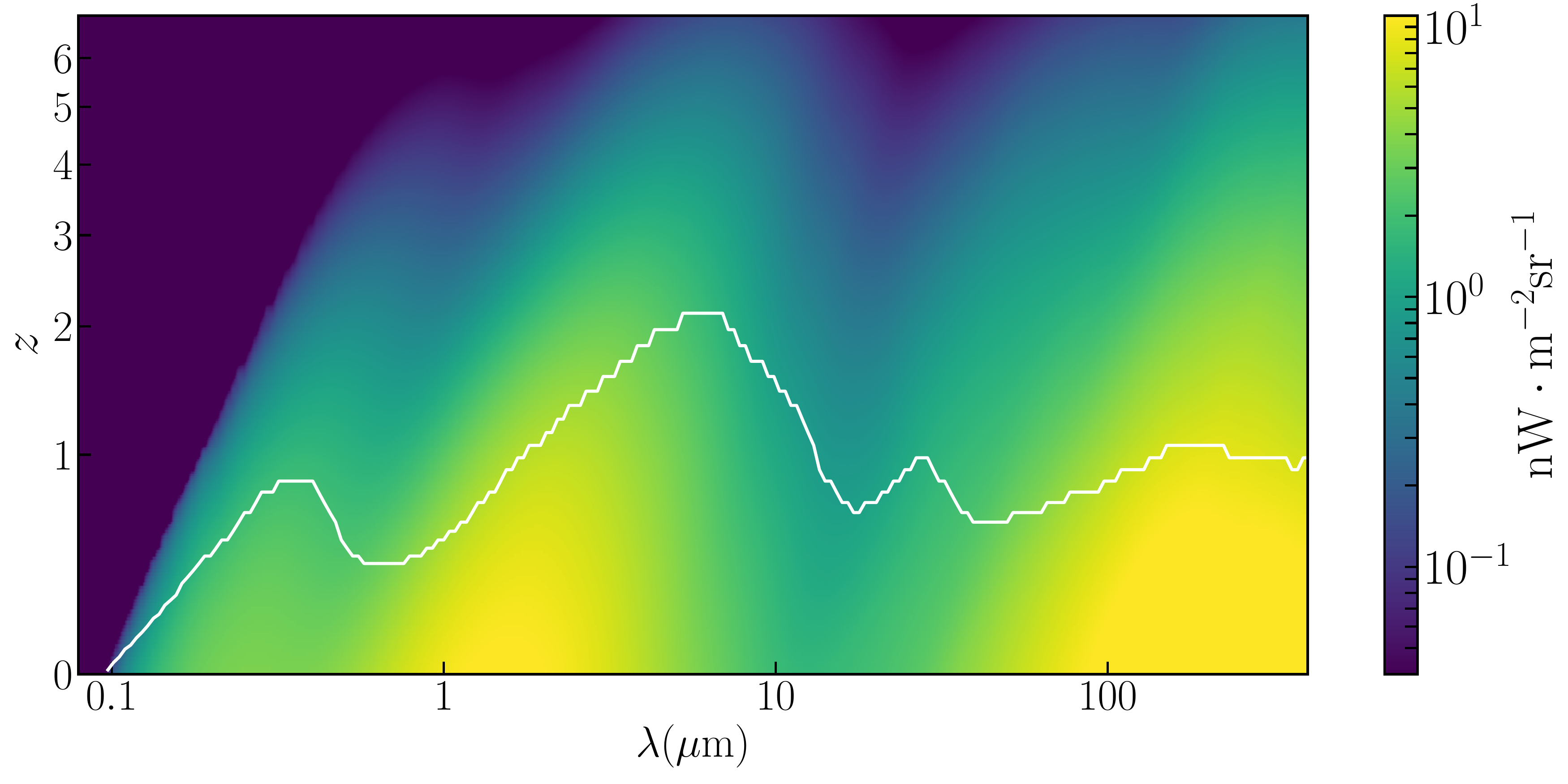}
  \end{center}
  \caption{ The build-up of the local EBL. The EBL at $z>0$ is shown in comoving coordinates as a function of the observed wavelength. {\it Left panel:} The median and 68\% confidence region of the build-up of the total background, integrated in the 0.1--8${\rm \mu m}$ range. Also displayed are $\gamma$-ray derived measurements of the COB at $z=0$ from \citet{hess_ebl13,biteau2015,Ahnen16} and integrated counts from \citet{driver16}. {\it Right panel:} The color map shows the median spectral intensity depicting the origin of the $z=0$ EBL at a given wavelength. The white continuous line marks the redshift at which 50\% of the local EBL has been accumulated. As the image shows the median EBL without uncertainties, we caution that the IR part is very poorly constrained i.e. towards the upper right corner of the image.
\label{fig:buildup}}
\end{figure*}

A key result of this work is a minimal EBL measurement throughout the wavelength range with respect to integrated galaxy light, allowing very little additional unresolved emission from faint or truly diffuse populations. We estimate the integrated cosmic background in the [0.09-8]\,{$\mu$m} range, often referred to as the cosmic optical background (COB), to be $27.8_{-2.0}^{+2.1}~{\rm nW\cdot m^{-2}sr^{-1}}$. We note that this is among the lowest { estimates} of the COB to date that is inferred from $\gamma$-ray data alone \citep[see also ][]{Ahnen16}. At the reference wavelength of 1.4${\rm \mu m}$ we find $\lambda I_\lambda = 11.8_{-1.23(2.2)}^{+2.2(5.2)}~{\rm nW\cdot m^{-2}sr^{-1}}$ ($1\sigma(2\sigma)$), limiting any undetected contribution to the cosmic near-IR background to $\lesssim 4~{\rm nW\cdot m^{-2}sr^{-1}}$ ($1\sigma$) with respect to integrated counts of \cite{driver16}, and even less with respect to \cite{keenan10}. This suggests that larger values of the EBL inferred by direct measurements that rely on absolute flux calibration are not extragalactic and likely attributable to Zodiacal light or other foreground emissions \citep{Matsuura17}.

An important aspect of this work is the ability to constrain the build-up of the EBL with cosmic time. This is illustrated in Figure \ref{fig:buildup} where we measure that 50\% of the COB has been accumulated by $z=0.9$. The build-up of the EBL across the entire wavelength range is qualitatively consistent with that of state-of-the-art EBL models \citep{Cowley2018}.

The fact that the reconstructed EBL is remarkably similar to integrated counts data, and a host of existing models, suggest that significant systematic biases in our analysis are unlikely. Known systematic uncertainties are already included in the optical depth uncertainties. With nearly $\sim 800$ blazars, any inaccuracies would need to affect the entire sample systematically in the same manner. Absorption intrinsic to the source \citep[largely ruled out now by][]{costamante18} for instance, from the black hole close environment or host galaxy, would result in the derived EBL being artificially larger, not lower. The fact that our EBL is already close to the minimum allowed by galaxy counts suggests that this effect, if present, is insignificant.

 Finally, our work makes use of latest blazar data from {\it Fermi}-LAT and present Cherenkov telescopes with a maximum energy of 21\,TeV and allows us to constrain the EBL up to 70\,$\mu$m. Long HAWC { observations} of bright blazars and observations by the upcoming CTA should be able to push this measurement even further, providing better IR constraints. At the same time, CTA should also be able to study the evolution of the EBL up-to a redshift of 0.5 with $10\%$ uncertainty \citep{cta17}, effectively complementing our results.


\section*{Acknowledgments}

A.Desai  and M.Ajello acknowledge funding support from NSF through grant AST-1715256. 
K.Helgason acknowledges support from the Icelandic Research Fund, grant
number 173728-051. The authors thank Dr.~Biteau for providing the TeV spectra used in this work in a machine readable format. A.Dom\'\iınguez thanks the support of the Ram{\'o}n y Cajal program from the Spanish MINECO.

\bibliographystyle{aasjournal}

\end{document}